\newcommand\AJ[3]{~Astron. J.{\bf ~#2}, #3~(#1)}
\newcommand\APJ[3]{~Astrophys. J.{\bf ~#2}, #3~(#1)}
\newcommand\apjl[3]{~Astrophys. J. Lett. {\bf ~#2}, L#3~(#1)}
\newcommand\ass[3]{~Astrophys. Space Sci.{\bf ~#2}, #3~(#1)}
\newcommand\cqg[3]{~Class. Quant. Grav.{\bf ~#2}, #3~(#1)}

\newcommand\mpla[3]{~Mod. Phys. Lett. A{\bf ~#2}, #3~(#1)}
\newcommand\npb[3]{~Nucl. Phys. B{\bf ~#2}, #3~(#1)}
\newcommand\plb[3]{~Phys. Lett. B{\bf ~#2}, #3~(#1)}
\newcommand\pr[3]{~Phys. Rev.{\bf ~#2}, #3~(#1)}
\newcommand\PRL[3]{~Phys. Rev. Lett.{\bf ~#2}, #3~(#1)}
\newcommand\PRD[3]{~Phys. Rev. D{\bf ~#2}, #3~(#1)}
\newcommand\prog[3]{~Prog. Theor. Phys. {\bf ~#2}, #3~(#1)}

\documentstyle[aps,preprint,floats,tighten]{revtex}
\input{epsfig.sty}
\newcommand\fracd[1]{\frac{#1'}{#1}}

\newcommand\VEV[1]{\left\langle #1 \right\rangle}
\begin{document}

\title{Cosmic Microwave Background Temperature and Polarization
Anisotropy in Brans-Dicke Cosmology}

\author{Xuelei Chen\thanks{Email: xuelei@phys.columbia.edu}
and Marc Kamionkowski\thanks{Email: kamion@phys.columbia.edu}}
\address{Department of Physics, Columbia University,\\
538 West 120th Street, New York, NY~~10027}

\date{May 27, 1999}
\preprint{CU-TP-942, CAL-686}
\maketitle

\begin{abstract}
We develop a formalism for calculating cosmic microwave background
(CMB) temperature and polarization anisotropies in cosmological
models with Brans-Dicke gravity.  We then modify publicly 
available Boltzmann codes to calculate numerically the
temperature and polarization power spectra.  Results are
illustrated with a few representative models. Comparing with the
general-relativistic model with the same cosmological
parameters, both the amplitude and the width of the acoustic
peaks are different in the Brans-Dicke models.  We use a
covariance-matrix calculation to investigate whether the effects 
of Brans-Dicke gravity are degenerate with those of variation in 
other cosmological parameters and to simultaneously determine
whether forthcoming CMB maps might be able to distinguish
Brans-Dicke and general-relativistic cosmology.  Although the
predicted power spectra for plausible Brans-Dicke models differ
{}from those in general relativity only slightly, we find that
MAP and/or the Planck Surveyor may in principle provide a test
of Brans-Dicke theory that is competitive to solar-system
tests. For example, if all other parameters except for the CMB
normalization are fixed, a value of the Brans-Dicke parameter
$\omega$ as large as 500 could be identified with MAP, and for
Planck, values as large as $\omega\simeq3000$ could be
identified; these sensitivities are decreased roughly by a
factor of 3 if we marginalize over the baryon density, Hubble
constant, spectral index, and reionization optical depth.  In
more general scalar-tensor theories, $\omega$
may evolve with time, and in this case, the CMB probe would be
complementary to that from solar-system tests.
\end{abstract}


\section{Introduction}
The Jordan-Fierz-Brans-Dicke theory 
\cite{Jordan,Fierz,Brans-Dicke} (heretofore, we will call
it Brans-Dicke theory for simplicity) is the simplest
example of a scalar-tensor theory of gravity
\cite{scalar-tensor}.  Recently, scalar-tensor
theories have received renewed interest, because such
theories are generic predictions of superstring theory \cite{superstring} 
and other higher-dimensional gravity theories \cite{Kaluza-Klein}.
Furthermore, scalar-tensor theories have also found application in
construction of inflationary models, including some models based 
on first-order phase transitions that evade the ``graceful
exit'' problem \cite{extended inflation,bubble,hyperextended
inflation,extended chaotic inflation}.

In Brans-Dicke theory, Newton's constant becomes a function of
space and time, and a new parameter $\omega$ is introduced.
General relativity is recovered in the limit $\omega\rightarrow
\infty$.  Solar-system experiments using Viking ranging data
\cite{Will} have constrained $\omega \geq 500$. 
Recent VLBI measurement of the time delay of millisecond 
pulsars may further raise this limit \cite{Will98}. 
However, these experiments are all
``weak-field'' experiments and probe only a limited range of space and
time.  To effectively constrain more general scalar-tensor
theories, one would  also like to have ``strong-field''
experiments, such as that provided by the binary
pulsar\cite{Will,binary-pulsar}.  It was also pointed out
\cite{attractor} that in cosmological models based on more general
scalar-tensor theories (in which $\omega$ can vary), there is generally 
an attractor mechanism that drives $\omega$ to $\infty$ at late 
times.  Thus, it is possible that gravity differed considerably
{}from general relativity in the early Universe, even if general
relativity seems to work well today.  Big-bang nucleosynthesis
\cite{BBN MK,BBN BD,BBN recent} provides one test of gravity at
early times.

With the advent of precise new cosmic microwave background (CMB) 
data, it is natural to inquire whether the CMB might be able to
provide new tests of Brans-Dicke theory (or of more general
scalar-tensor theories). The Microwave Anisotropy Probe (MAP) 
\cite{MAP} (to be launched in Fall 2000) and 
Planck Surveyor \cite{Planck} (to be launched in 
2007) satellites as well as many ground-based and balloon-borne experiments 
will measure the CMB anisotropy with unprecedented precision, 
thus providing a wealth of information about the early Universe.
The advantage of the CMB anisotropy is that it involves fairly
simple linear physics and is thus very ``clean''. Furthermore,
the CMB anisotropy probes a different era of the cosmos.
Thus, at least in principle one may see the presence of a
scalar-tensor theory that has been driven by the attractor mechanism 
to the general-relativity limit in the current epoch
\cite{attractor}.

The possibility of testing scalar-tensor gravity with CMB anisotropy
has already been noted. For example, the original version of
extended inflation \cite{extended inflation} was ruled out because the
bubbles formed during the phase transition would have produced
CMB anisotropy larger than that observed unless the Brans-Dicke
parameter $\omega$  was less than 30 \cite{bubble}.

The general behavior of cosmological perturbations in
Brans-Dicke cosmology was studied analytically in
Refs. \cite{Nariai,Baptista}, but they did not consider  
realistic models.  Peebles and Yu, in their pioneering study of 
CMB anisotropy \cite{Peebles and Yu}, considered a more realistic
model with Brans-Dicke gravity, and they showed how the
difference in the expansion rate affects the photon transfer
function.  More recently, Liddle et al. \cite{Liddle} estimated
that in Brans-Dicke theory, the epoch of radiation-matter
equality is shifted,
\begin{equation}
\frac{a_{\rm eq} H_{\rm eq}}{a_{0} H_{0}} = 219 h
\left(1+\frac{5.81}{\omega}+ \frac{\ln h}{\omega}\right),
\end{equation}
and this accordingly affects the scale at which the present-day
matter power  spectrum turns over.

In the particular case of cosmologies based on chaotic-inflation
models, the production of fluctuations during inflation
with scalar-tensor
gravity has been studied in
Refs. \cite{Starobinsky-Yokoyama,Garcia-Bellido-Wands,Chiba et
al}. They concluded that isocurvature perturbations could be
produced during inflation, but are in general negligible
compared with the adiabatic perturbations.  In some inflation
models, the spectrum of density perturbations may be affected,
and for scalar-tensor theories with variable $\omega$, the
spectral index for primordial perturbations may change with
scale. For example, in some Brans-Dicke inflationary models 
there is a slight tilt in the spectrum of density perturbations,
and a limit on
the variation of $\omega$ can be obtained from the COBE
observation measurement of the spectral index \cite{COBE-DMR},
but only within the context of this very particular inflation model.

In this paper, we perform a complete calculation of the CMB
anisotropy in the Brans-Dicke theory. To do this we modify a
standard code for CMB anisotropy calculation \cite{CMBFAST}
to accommodate Brans-Dicke theory.  Our modified code may be
used to calculate the anisotropy in any given cosmological model.
Although our code can accommodate isocurvature perturbations as
well, we present numerical results only for models with nearly
scale-invariant spectra of primordial adiabatic perturbations
for the following reasons:  If acoustic peaks like those
expected from adiabatic perturbations are observed, then it is
plausible that we might understand structure formation well
enough to use CMB anisotropy to look for tiny deviations from
general relativity.  If it appears that some more complicated
physics gave rise to structure formation, then it is unlikely
that the CMB will provide a precision tool to study gravity.

We limit ourselves here to the simplest scalar-tensor theory:
i.e., the original Brans-Dicke theory, for which the
Brans-Dicke parameter $\omega$ is fixed.  We will leave the more general
case with variable $\omega$ to future work.  Likewise, we
concentrate on flat CDM models, including those with a
cosmological constant, but without hot dark matter.

This paper is organized as follows: In Section II, we develop the
formalism for the calculation.  In Section III, we study the
behavior of the background cosmology and discuss the initial
conditions for the perturbations in the Brans-Dicke field.
Numerical results are presented in Section IV, and we also
discuss the detectability of Brans-Dicke theory there.
Section V then summarizes and concludes.  We briefly describe the
numerical implementation of the calculation in the Appendix.
Throughout this paper, we use natural units, $c=\hbar=k_{B}=1$.

\section{Formalism}

\subsection{Brans-Dicke Theory}

The Lagrangian density for the Brans-Dicke theory is 
\begin{equation}
\label{lagrangian}
{\mathcal L}=\sqrt{-g}\left[-\Phi R+
\frac{\omega}{\Phi}g^{\mu\nu}\partial_{\mu}\Phi
\partial_{\nu}\Phi+L_{m}\right],
\end{equation}
where $\Phi$ is the Brans-Dicke field, and $L_{m}$ is the
Lagrangian density for the matter fields, whose equations of
motion are not affected.  For convenience, we also define a
dimensionless field
\begin{equation}
\phi=G \Phi,
\end{equation}
where $G$ is the Newtonian gravitational constant measured today.

The Einstein equations are generalized to
\begin{eqnarray}
G_{\mu\nu}&=&\frac{8\pi}{\Phi}T_{\mu\nu} + \frac{\omega}{\Phi^2}
(\Phi_{;\mu} \Phi_{;\nu} -\frac{1}{2}g_{\mu\nu}
\Phi_{;\lambda}^{\;\;;\lambda})
+\frac{1}{\Phi} (\Phi_{;\mu\nu}-g_{\mu\nu}\Box \Phi),
\end{eqnarray}
where $T_{\mu\nu}$ is the stress tensor for all matter except for the 
Brans-Dicke field. The equation of motion for $\Phi$ is 
\begin{equation}
\Box\Phi= \frac{8\pi}{2\omega+3} T.
\end{equation}
Here $T=T^{\mu}_{\;\mu}$ is the trace of the energy-momentum
tensor.

\subsection{Background Cosmology}

The unperturbed part of the metric in a flat universe can be written as
\begin{equation}
ds^2 = a^{2}(\tau)(-d\tau^{2} + \gamma_{ij} dx^i dx^j ).
\end{equation}
where $a$ is a function of the conformal time $\tau$ only, and
$\gamma_{ij}$ is the flat-space metric. The unperturbed
stress-energy tensor has components
\begin{equation}
T^{0}_{\;0}=-\rho, \qquad T^{0}_{\;i}=0, \qquad T^{i}_{\;j}=p
\gamma^{i}_{\;j},
\end{equation}
in the comoving frame. The equations describing the background evolution are  
\begin{equation}
\label{Heq}
\rho' + 3\fracd{a} (\rho+p)=0,
\end{equation}
\begin{equation}
\label{modified FRW}
\left(\fracd{a}\right)^2 = \frac{8\pi G a^2}{3\phi} \rho
+\frac{\omega}{6}\left(\fracd{\phi}\right)^2
-\fracd{a}\fracd{\phi},
\end{equation}
\begin{equation}
\phi''+2\fracd{a}\phi'=\frac{8\pi G a^2 }{2\omega+3} (\rho-3p),
\label{phieq}
\end{equation}
where the prime denotes derivative with respect to $\tau$, and 
$\rho$ and $p$ are the total density and pressure of the
Universe, respectively.  General relativity is recovered in the
limits
\begin{equation}
\Phi'' \to 0, \quad \Phi' \to 0, \quad \omega \to \infty.
\end{equation}

\subsection{Cosmological Perturbations}

We can write the perturbed metric as
\begin{equation}
g_{\mu\nu}=a^{2} (\gamma_{\mu\nu}+h_{\mu\nu}),
\end{equation}
where the perturbation $h_{\mu\nu}$ is a function of space and
time.  It will be easier for us to deal with
the (spatial) Fourier components of $h_{\mu\nu}$, and to avoid cluttered
notation, we will subsequently denote the Fourier components
$\tilde{h}_{\mu\nu}({\bf k})$ simply by $h_{\mu\nu}$.
We choose to work in synchronous gauge, so
$h_{00}=h_{0i}=h_{i0}=0$, and the $h_{ij}$ can be expanded in
tensor harmonics, which satisfy $\nabla^{2}
{\mathbf Q}^{(m)} = -k^{2} {\mathbf Q}^{(m)}$
\cite{Kodama-Sasaki,Hu et al},
\begin{eqnarray}
h_{ij} &=& \sum_{m} 2H^{(m)}_{L} Q^{(m)} \gamma_{ij} + 2 H^{(m)}_{T} 
Q^{(m)}_{ij},\\
\delta\phi&=& \sum_{m} \chi^{(m)} Q^{(m)},
\end{eqnarray}
where $Q^{(m)}$ and $Q^{(m)}_{ij}$ are scalar and tensor harmonics,
respectively, and $m$ denotes the ``angular momentum'' of the
perturbation.  For simplicity, we will write
\begin{eqnarray}
H_{L}^{(0)}&=&h_{L}, \qquad H_{T}^{(0)}=h_{T},\nonumber\\
H_{T}^{(1)}&=&h_{V}, \qquad H_{T}^{2}=H.
\end{eqnarray}
For models with only scalar and tensor modes, $h_{V}=0$. Our
$h_{L}$ and $h_{T}$ are simply related to the variables used in
Ref. \cite{Ma-Bertschinger} by
\begin{equation}
h = 6 h_{L}, \qquad \eta = -(h_{L}+h_{T}/3).
\end{equation}

The perturbed stress energy tensor can also be broken up into scalar,
vector, and tensor parts. Let us denote a cosmic fluid component
(e.g., baryons, neutrinos, photons, cold dark matter, etc.) by
index $f$.  We then know
that the stress-energy perturbations $\delta T_i^{\;j}$ are
related to the perturbations $\delta\rho$ and $\delta p$ in the
density and pressure, respectively, and to the velocities $v_f$
and anisotropic stress $\pi_f$ (see, e.g., 
Refs.~\cite{Ma-Bertschinger,Hu et al} for more details),
\begin{eqnarray}
\delta T^{0}_{\;0}&=&-\sum_f \sum_{m}\delta\rho^{(m)}_{f} Q^{(m)},\nonumber\\
\delta T^{0}_{\;i}&=&\sum_f \sum_{m} (\rho_{f}+p_{f}) v_{f}^{(m)} 
Q_{i}^{(m)},\nonumber\\
\delta T_{0}^{\;i}&=&-\sum_f \sum_{m} (\rho_{f}+p_{f}) v_{f}^{(m)} Q^{(m)i},
\nonumber\\
\delta T^{i}_{\;j}&=&\sum_f \sum_{m}\delta p_{f}^{(m)}
\gamma^{i}_{\;j}Q^{(m)}+p_{f}\pi_{f}^{(m)} Q^{(m)i}_{j}.
\end{eqnarray}
If we consider only scalar and tensor perturbations, then the
perturbed Einstein and Brans-Dicke equations are
\begin{eqnarray}
\label{perturbed}
\chi''+2\fracd{a}\chi'+k^2 \chi+3h'_{L}\phi' &=& \frac{8\pi G a^2}{2\omega+3}
\sum_{f}\left(\delta\rho_{f}-3\delta p_{f}\right); 
\end{eqnarray}
\begin{eqnarray}
\label{forcecon}
k^2 \left(h_{L}+\frac{1}{3}h_{T}\right)+3\fracd{a}h_{L}'&=&
\frac{4\pi G a^2}{\phi}\sum_{f}\rho_{f}\delta_{f}
-\frac{3}{2}(\fracd{a})^{2}\frac{\chi}{\phi}
-\frac{\omega{\phi'}^2\chi}{4\phi^3}
+\frac{\omega\phi'\chi'}{2\phi^{2}}\nonumber\\
&&-\frac{1}{2} \left[k^2 \frac{\chi}{\phi}
+3 h'_{L}\fracd{\phi}+3\fracd{a}\frac{\chi}{\phi}\right];
\end{eqnarray}
\begin{eqnarray}
h_{L}'+\frac{1}{3}h_{T}' &=& -\frac{4\pi G a^{2}}{\phi}\sum_{f}(\rho_{f}+p_{f})
\frac{v_{f}}{k} -\frac{\omega\phi'\chi}{2\phi^{2}}
-\frac{1}{\phi}
\left(\chi'-\fracd{a}\chi\right);
\end{eqnarray}
\begin{eqnarray}
h_{T}''+2 \fracd{a} h_{T}'-k^2 (h_{L}+\frac{1}{3}h_{T})&=&
\frac{8\pi a^{2}}{\phi} p\pi_{f} -h_{T}'\fracd{\phi}+k^{2} \frac{\chi}{\phi};
\end{eqnarray}
\begin{equation}
H''+2\fracd{a}H'+k^{2} H=\frac{8\pi G a^{2}}{\phi} p_{f} \pi_{f}^{(2)}.
\label{perturbed-last}
\end{equation}

\subsection{Temperature and Polarization Anisotropies}

With these equations, one can find the evolution of
perturbations using standard cosmological perturbation 
theory; see e.g. Ref. \cite{Kodama-Sasaki}.  The calculation of the CMB
anisotropy runs in parallel to the one in the standard model
detailed, e.g., in
Refs. \cite{Ma-Bertschinger,Seljak-Zaldarriaga,Hu et al}.
Here we summarize the procedure for such calculations.

A temperature map $T(\hat{\mathbf n})$ of the sky (as a function of
position $\hat{\mathbf n}$ on the sky) can be expanded in spherical
harmonics,
\begin{equation}
     {T(\hat{\mathbf n}) \over T_0} = 1 + \sum_{lm} a^{\rm T}_{(lm)}
     Y_{(lm)}(\hat{\mathbf n}),
\label{Texpansion}
\end{equation}
where the mode amplitudes are given by the inverse
spherical-harmonic transform.
Similarly, if we measure the Stokes parameters $Q(\hat{\mathbf n})$ and
$U(\hat{\mathbf n})$ as a function of position on the sky, they can be
assembled into a symmetric trace-free (STF) $2\times2$ tensor
\cite{CMB polarization KKS},
\begin{equation}
  {\cal P}_{ab}(\hat{\mathbf n})={1\over 2} \left( \begin{array}{cc}
   \vphantom{1\over 2}Q(\hat{\mathbf n}) & -U(\hat{\mathbf n}) \sin\theta \\
   -U(\hat{\mathbf n})\sin\theta & -Q(\hat{\mathbf n})\sin^2\theta \\
   \end{array} \right),
\label{whatPis}
\end{equation}
which can then be be expanded \cite{CMB polarization KKS},
\begin{equation}
      {{\cal P}_{ab}(\hat{\mathbf n})\over T_0} =
      \sum_{lm} \left[ a_{(lm)}^{{\rm G}}Y_{(lm)ab}^{{\rm
      G}}(\hat{\mathbf n}) +a_{(lm)}^{{\rm C}} Y_{(lm)ab}^{{\rm
      C}}(\hat{\mathbf n})
      \right],
\label{Pexpansion}
\end{equation}
where the tensor spherical harmonics $Y_{(lm)ab}^{\rm G}$ and
$Y_{(lm)ab}^{\rm C}$  form a complete basis for the gradient and
curl components of the tensor field, respectively, and the
multipole coefficients, $a_{(lm)}^{{\rm G}}$ and $a_{(lm)}^{{\rm
C}}$ can be obtained by inverse transforms.

Thus, a combined temperature/polarization map is specified
completely by the three sets of coefficients, $a_{(lm)}^{\rm
T}$, $a_{(lm)}^{\rm G}$, and $a_{(lm)}^{\rm C}$.  The two-point
statistics of the T/P map are specified completely by the six
power spectra $C_l^{{\rm X}{\rm X}'}$ defined by
\begin{equation}
     \VEV{\left(a_{(l'm')}^{\rm X'} \right)^* a_{(lm)}^{\rm X}} =
     C_l^{{\rm XX}'} \delta_{ll'}\delta_{mm'},
\end{equation}
for ${\rm X},{\rm X}' = \{{\rm T,G,C}\}$, and the angle brackets 
denote an ensemble average.  Parity invariance
demands that $C_l^{\rm TC}=C_l^{\rm GC}=0$.  Therefore the
statistics of the CMB temperature-polarization map are
completely specified by the four sets of moments, $C_l^{\rm
TT}$, $C_l^{\rm TG}$, $C_l^{\rm GG}$, and $C_l^{\rm CC}$.
These correlation functions are related to the ones used
by Seljak and Zaldarriaga \cite{CMB polarization SZ} by
$C^{\rm GG}_{l}=C_{El}/2$, $C^{\rm CC}_{l}=C_{Cl}/2$, $C^{\rm
TG}_{l}=C_{Cl}/\sqrt{2}$, and our $C^{TT}_{l}$ is the same as
their $C_{Tl}$.

We can calculate the $C_{l}$'s by convolving the initial metric perturbation
power spectrum $P_{\psi}$ with the photon transfer function 
$\Delta_{l}(k, \tau_{0})$,
\begin{eqnarray}
\label{losint1}
C_l^{\rm T}&=&(4\pi)^2 \int k^2 dk P_{\psi}(k) 
[\Delta_{Tl}(k)]^2,\\
C_{l}^{\rm GG}&=&(4\pi)^2 \int k^2 dk P_{\psi}(k) 
[\Delta_{Gl}(k)]^2,\\
C_l^{\rm TG}&=&(4\pi)^2 \int k^2 dk P_{\psi}(k) 
[\Delta_{Tl}\Delta^{(S)}_{Gl}].
\label{losint2}
\end{eqnarray}
The photon transfer functions $\Delta_{Xl}(k, \tau_{0})$ are obtained by
integrating along the line of sight \cite{Seljak-Zaldarriaga},
\begin{eqnarray}
\Delta_{Tl}(k, \tau_{0})&=&\int_{0}^{\tau_{0}} d\tau
S_{T} (k, \tau) j_{l}[k(\tau_{0}-\tau)],\\
\Delta_{Gl}(k, \tau_{0})&=&\sqrt{\frac{(l+2)!}{(l-2)!}}
\int_{0}^{\tau_{0}} d\tau S_{G}(k, \tau) j_{l}[k(\tau_{0}-\tau)],
\end{eqnarray}
where $j_{l}(x)$ is the spherical Bessel function and $S_{T,G}(k, \tau)$ are
the source functions describing the Thomson scattering of photons along
the path, and
\begin{eqnarray}
S_{T}&=&g\left(\Delta_{T0}+2\alpha'+\frac{v_{b}'}{k}
+\frac{\Pi}{4}+\frac{3\Pi''}{4k^{2}}\right)+e^{-\kappa}(\eta'+
\alpha'')\nonumber\\
&&+ g'\left(\frac{v_{b}}{k}+\frac{3\Pi'}{4k^{2}}\right)
+\frac{3g''\Pi}{4k^{2}},\\
S_{G}&=&\frac{3 g(\tau) \Pi(\tau,k)}{8 (k\tau)^2 },\\
\Pi&=&\Delta_{T2}+\Delta_{G2}+\Delta_{G0},\nonumber
\end{eqnarray}
where $x=k(\tau_{0}-\tau)$ and $\alpha=(h'+6\eta')/2k^{2}$. The 
visibility function $g(\tau)$ is given by $g=e^{-\kappa}
\kappa'$, where $\kappa(\tau)$ is the optical depth from conformal
time $\tau$ to the current epoch.
In the Brans-Dicke theory, the derivatives of $\alpha$ are given by
\begin{eqnarray}
\label{alpha1}
\alpha'&=&-\frac{1}{k^{2}}\fracd{a}\left(h'+6\eta'\right)+\eta
-\frac{8\pi G a^2}{\phi} \frac{p \pi_{f}}{k^2} 
+\frac{1}{2k^2}\left(h'+6\eta'\right)\fracd{\phi}-\frac{\chi}{\phi},\\
\alpha''&=&-2\left(\fracd{a}\right)'\alpha
-2\left(\fracd{a}\right)\alpha'+\eta'
-\frac{3}{2k^2}\frac{8\pi G}{\phi}\left[a^2
\left(\rho+p\right)\sigma\right]'\nonumber\\
&&+\frac{3}{2k^2 }\frac{8\pi G a^2}{\phi}\left(\rho+p\right)\sigma
\fracd{\phi}-\alpha'\fracd{\phi}+\alpha\left(\fracd{\phi}\right)'
-\left(\frac{\chi}{\phi}\right)'.
\label{alpha2}
\end{eqnarray}

For the initial conditions on the scalar-field perturbation, we
consider only the simplest case with $\chi_{\rm init}=\chi'_{\rm
init}=0$.  The initial conditions for the matter are the same as
those in the GR case \cite{Ma-Bertschinger}.
As perturbations in the metric grow, perturbations in the Brans-Dicke
field will also grow as shown in Eq.~(\ref{perturbed}). However,
for the initial condition we have chosen, the Brans-Dicke
perturbation is so small that it has little effect on CMB
anisotropy.  An alternative choice of initial conditions for the 
Brans-Dicke perturbation would probably yield the same numerical 
results, because any initial perturbations are damped during the
radiation dominated era [c.f. Eq.~(\ref{perturbed})].

The numerical calculation is essentially carried out by
replacing the general-relativistic perturbation equations in a
publicly available code \cite{CMBFAST,COSMICS} by those in
Eqs. (\ref{forcecon})--(\ref{perturbed-last}) and including
the evolution equation (\ref{perturbed}) for the Brans-Dicke field.
In practice, there are a number of numerical issues and
subtleties that arise, and some of these are detailed in the
Appendix.

We have chosen to work in the Jordan frame in which the
equations for spacetime-metric perturbations are altered while
the equations for the stress-energy perturbations are
unchanged.  We considered working in the Einstein frame, in
which the metric-perturbation equations are unchanged, but found
that the changes in the equations for the stress-energy tensor
would be more difficult to implement numerically.

\section{Background Cosmology}

Let us now consider the background cosmology, and the boundary
conditions for the homogeneous component of the Brans-Dicke
field $\Phi$ and its conformal-time derivative $\Phi'$. We
define the  cosmic scale factor at the present epoch to be $a_0=1$.
In general-relativistic cosmology, the initial condition for the 
scale factor is $a(\tau=0)=0$.  The conformal age of the
Universe can be obtained by integrating,
\begin{equation}
\label{daoa}
\int_{0}^{\tau_{0}} d\tau= \int_{0}^{1} da/a'.
\end{equation} 

For Brans-Dicke cosmology,
additional boundary conditions are required for $\phi$ and $\phi'$.
One of these is determined by the requirement that the gravitational
constant be in agreement with that measured today. This fixes \cite{Brans-Dicke} 
\begin{equation}
\phi= \frac{2\omega+4}{2\omega+3}
\label{init1}
\end{equation}
at the present epoch.

The cosmological solutions for Brans-Dicke theory have been studied 
extensively \cite{Nariai,Gurevich,Barrow-Parsons,Holden-Wands}.
The Brans-Dicke field has a stiff
equation of state; it dominates the dynamics 
at early stages of the expansion. However, 
for the era which affects the CMB
anisotropy, the Brans-Dicke field must be subdominant, or else
the expansion rate at nucleosynthesis would have been very
different. Therefore, for a qualitative understanding of the
expansion, we can assume that the change in
$\phi$ does not affect the dynamics, and estimate how $\phi$ varies by
assuming the Universe expands as in the GR case.

\begin{figure}[htbp]
\begin{center}
\epsfig{file=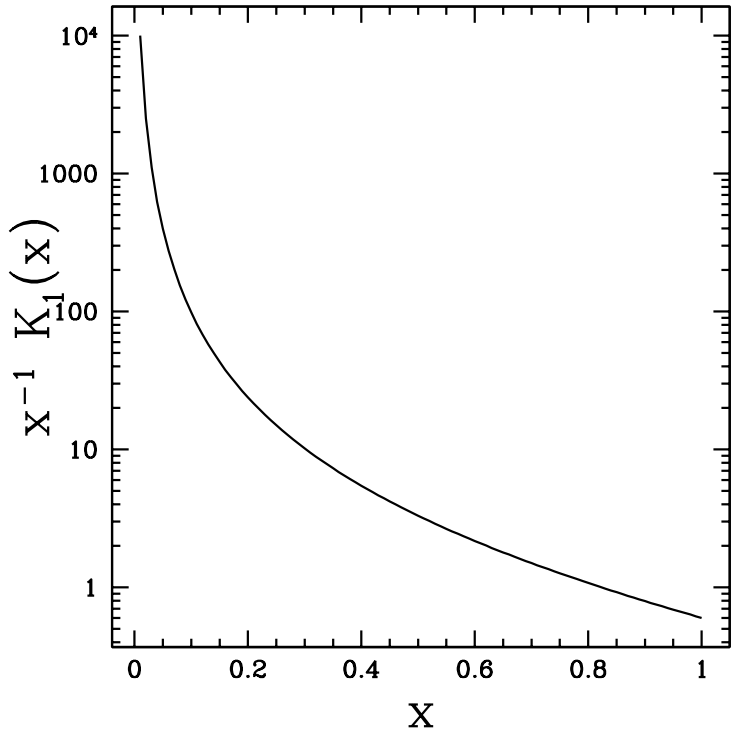,width=4in}
\caption{The function $K_{1}(x)/x$ versus $x=m/T$ (increasing time).}
\label{fig:BesselK1}
\end{center}
\end{figure}

The equation of motion for the $\phi$ field is given by
Eq. (\ref{phieq}).  It is analogous to a damped oscillator with 
a variable friction force. The initial ``velocity'' $\phi'$
is damped in a few Hubble times.  Therefore, for most of the
time concerned, $\phi$ would only vary slowly.
The right-hand side of Eq. (\ref{phieq}) is the ``driving
force'' for the motion of
$\phi$. It is proportional to $\rho-3p$.  
If $\rho-3p$ were to vanish during
radiation domination, then the quantity
$y\equiv a^2 \phi'$  would be constant. However, it does not vanish
during the radiation dominated era; in fact it is greater, even
though the ratio $(\rho-3p)/\rho$ is smaller. There are two kinds of
contributions to $\rho-3p$.  First, the non-relativistic matter,
including both baryons and cold dark matter, always
contributes.  Second, if one massive relativistic particle is
present, it always dominates.\footnote{Here we do not consider the 
very early era during which everything, including baryons and
cold-dark-matter particles were still relativistic.}
For one species of massive relativistic particles,  
\begin{equation}
\rho-3p=\frac{g}{2\pi^{2}}\int_{0}^{\infty} \left(E(p)-
\frac{p^{2}}{E(p)}\right) p^{2} f(p) dp,
\end{equation} 
where $f(p)$ is the distribution function. For example, for a 
Boltzmann gas with zero
chemical potential, $f(p)=e^{-E/T}$, and the result can be 
expressed in modified Bessel functions,
\begin{equation}
\rho-3p=\frac{g}{2\pi^2 m^4} \frac{K_{1}(m/T)}{m/T}.
\end{equation}
Fig.~\ref{fig:BesselK1} plots this function.  As one can see,
as $T\to \infty$, $m/T \to 0$, and this function rises rapidly.

In the present paper, we will not consider massive neutrinos. In a
CDM model, the last decoupled massive relativistic particles are electrons
and positrons. They annihilate below \hbox{$T\approx m_{e}= 0.511$ MeV}.
For the scale we are interested in, the main contribution comes from
the cold dark matter, which scales as $\rho_{c}=\rho_{c0}a^{-3}$.

After $e^{+}e^{-}$ annihilation, in the radiation dominated era, 
\begin{equation}
a \propto \tau, \quad \phi'\approx c_{1}+c_{2}a^{-2},
\end{equation}
and one can see that $\phi$ approaches a ``terminal velocity'' and the
initial velocity quickly dies out. From $T\sim 0.5$ MeV to
$T\sim 10$ eV (matter-radiation equality), the initial velocity of
$\phi$ is suppressed by a factor of $10^{9}$, and this initial
``velocity'' is constrained by nucleosynthesis, so it cannot
be too large. We estimate,
\begin{equation}
\frac{2\omega+3}{12}\left(\fracd{\phi}\right)^2 <
\left(\fracd{a}\right)^2 ,
\end{equation}
at the end of nucleosynthesis. We find that for all practical
purposes, we can take 
\begin{equation}
a^2 \phi_{i}'=0.
\label{init2} 
\end{equation}
This is effectively the Brans-Dicke initial condition
\cite{Brans-Dicke} proposed in their first paper.

\begin{figure}[htbp]
\begin{center}
\epsfig{file=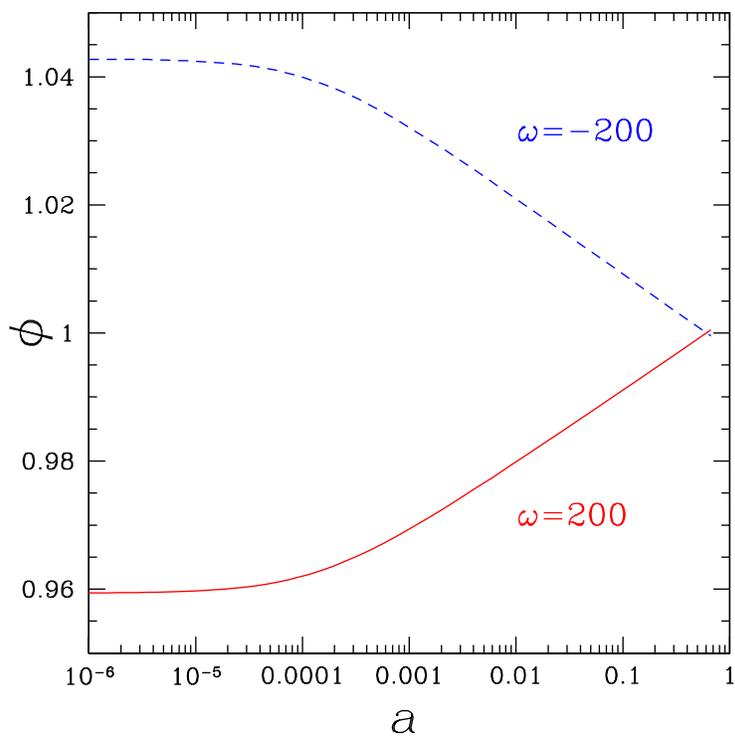,width=4in}
\caption{The time evolution of the Brans-Dicke field $\phi$.}
\label{fig:phievol}
\end{center}
\end{figure}

In the matter-dominated era, $\phi$ varies as
 $\phi \propto a^{1/(\omega+1)}$. For models with $\omega > 0$, the
value of $\phi$ increases with time, whereas for models with
$\omega<0$, $\phi$ decreases with time.  Fig.~\ref{fig:phievol}
shows the evolution of $\phi$.

We also note that in Brans-Dicke theory, the matter density is
not precisely equal to the critical density in a flat Universe.
The critical density in the Brans-Dicke theory depends on the
parameter $\omega$. If we still define our relative densities in
the usual way, i.e.,
\begin{equation}
\Omega_i=\rho_i/\rho_{c}, \quad \rho_{c} = 3 H_{0}^2 /8\pi G, 
\end{equation}
then we will have $\Omega=\sum_{i}\Omega_{i}\neq 1$ for flat
geometry. Let us define 
\begin{equation}
D\equiv \left(\fracd{\phi}\right)_{a=1};
\end{equation}
then from Eq. (\ref{Heq}) we have
\begin{equation}
\frac{\Omega}{\phi}=1+\frac{D}{H_{0}}
-\frac{\omega}{6} \left(\frac{D}{H_{0}}\right)^{2}.
\label{eq:changedOmega}
\end{equation}
With the matching condition, $\phi=(2\omega+4)/(2\omega+3)$, one can
obtain the value of $\Omega$ if $D$ is known. To proceed, one may
start with some value of $\rho$ and solve the evolution
equation to obtain $H_{0}$ and $D$.  In practice, for the models
considered here the difference is very small. We have tested for a few
models with $\omega$ equal to a few hundred, and it does not
make any significant difference.

\begin{figure}[htbp]
\begin{center}
\epsfig{file=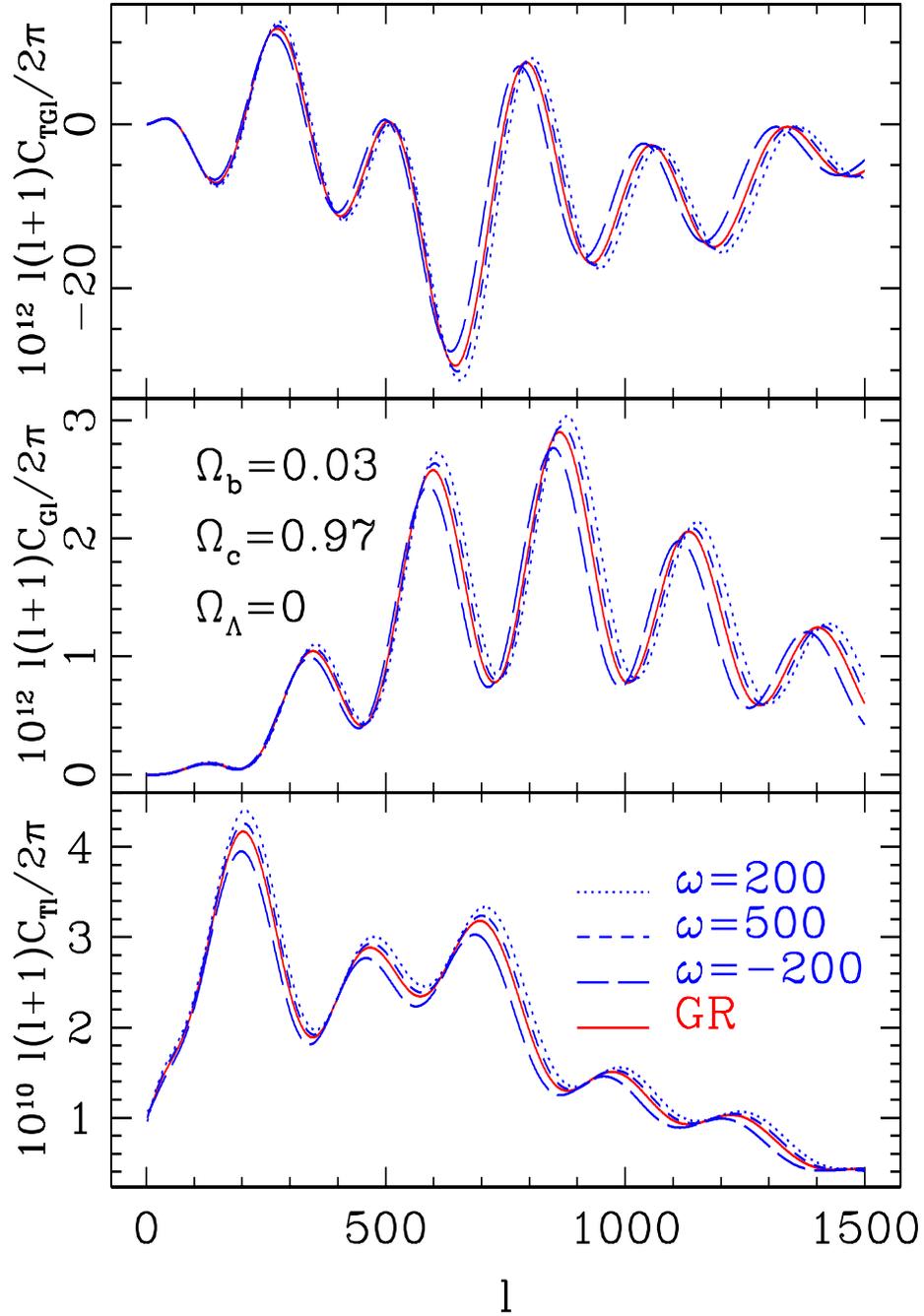,width=5in,height=7.5in}
\caption{CMB temperature/polarization power spectra for flat SCDM
models in general relativity and in Brans-Dicke theories with
$\omega=200$, $500$, and $-200$.}
\label{fig:combined}
\end{center}
\end{figure}

\section{Results}

We now illustrate our numerical results with a few representative 
models.  First we consider a
COBE-normalized flat standard CDM (SCDM) model, with $\Omega_{b}=0.03$,
$\Omega_{c}=0.97$ and $h=0.65$.
The angular power spectra for $\omega=200$, 500, and $-200$ 
are plotted in 
Fig. \ref{fig:combined} For comparison, we have also plotted the
general-relativity result with the same physical parameters in
the same plot. 

For $\omega=\pm 200$ the difference between Brans-Dicke models 
and general relativity are clearly discernible
As can be seen,
both the normalization and width of the acoustic peaks are changed. 
The Brans-Dicke model with a positive $\omega$ 
has higher and broader acoustic peaks,
while the negative-$\omega$ model has lower and narrower peaks.
We have checked that the perturbations in the Brans-Dicke field
near the time of decoupling in this model are very small.  Thus, 
change in the acoustic-peak structure is due  primarily to the
change in the expansion rate near decoupling. 
For $\omega=500$ 
the difference is much less pronounced.
The polarization spectra are similarly affected.

\begin{figure}[htbp]
\begin{center}
\epsfig{file=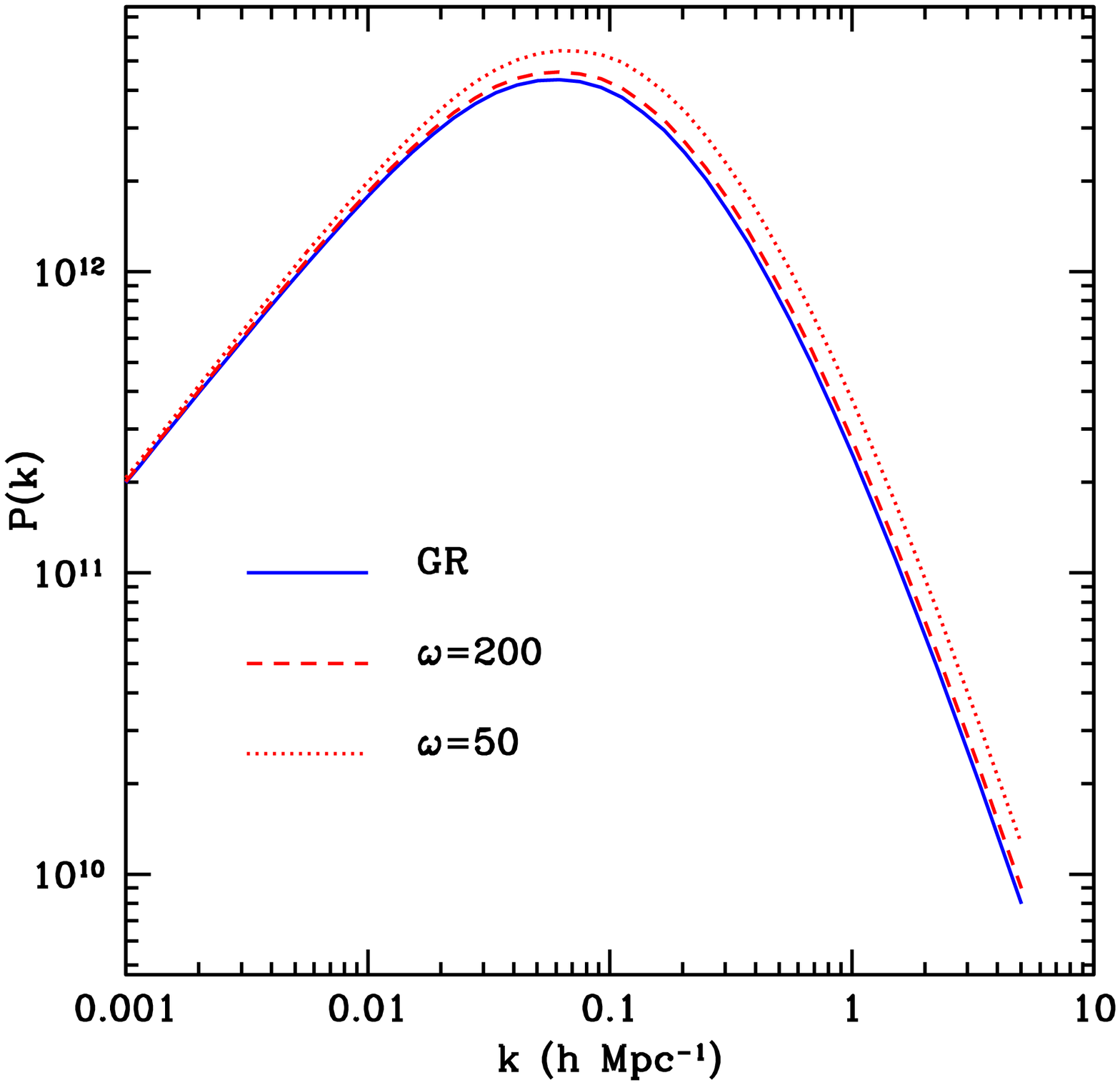,width=4in}
\caption{Spatial power spectra for SCDM models in Brans-Dicke theory with
   $\omega=200$ and $\omega=50$ and for general relativity.}
\label{tfplot}
\end{center}
\end{figure}

The Brans-Dicke field also affects the transfer
function. Fig.~\ref{tfplot} compares the matter transfer function in a
general-relativity model, a Brans-Dicke model  with $\omega=200$, and one
with $\omega=50$. For the Brans-Dicke models, the bend of the matter 
power spectrum occurs at shorter wavelengths, and there is thus
more small-scale power, in agreement with the claims of
Ref. \cite{Liddle}.

\begin{figure}[htbp]
\begin{center}
\epsfig{file=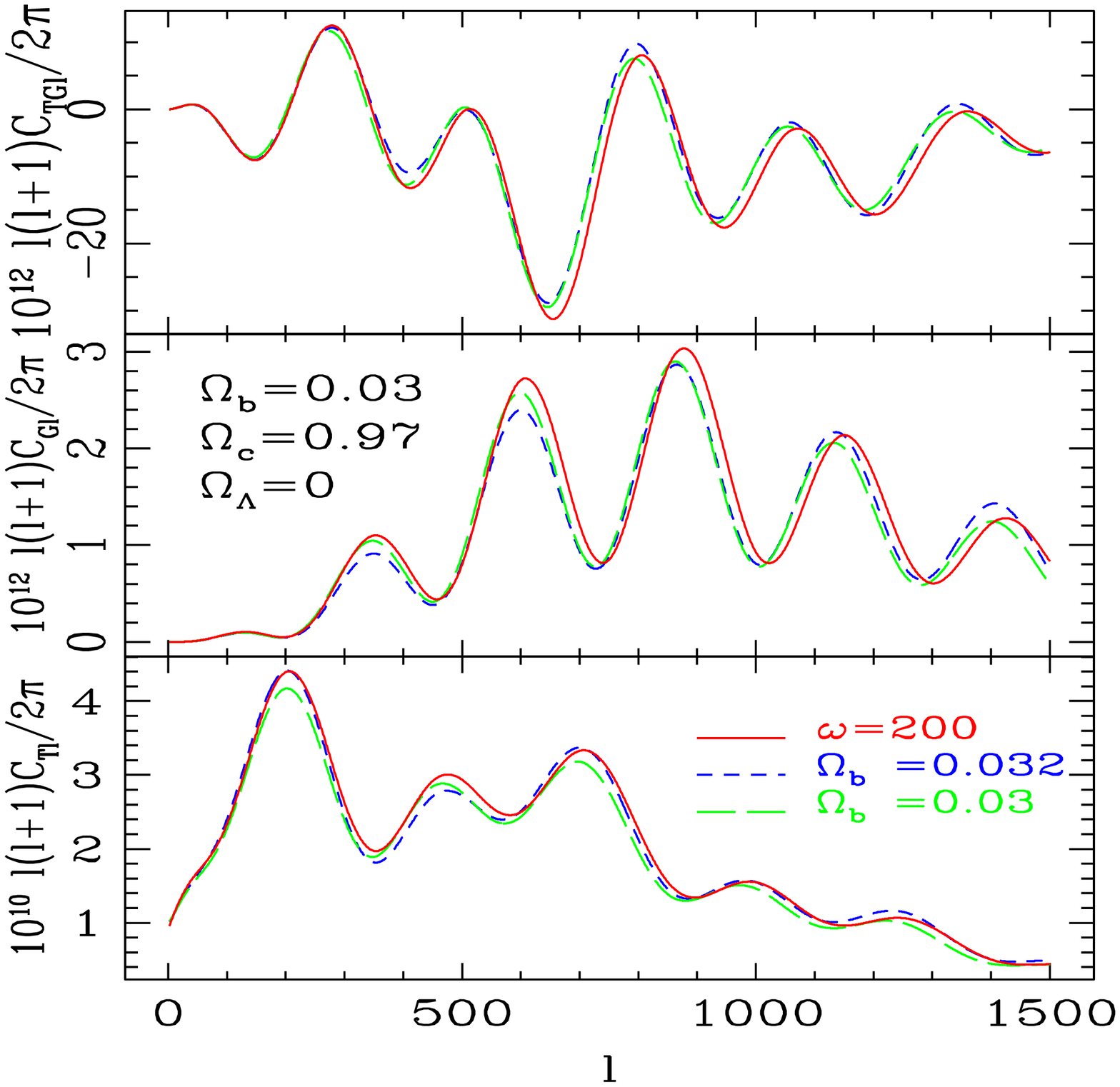,width=5in,height=7.5in}
\caption{CMB power spectra for a Brans-Dicke SCDM model with
     $\Omega_b=0.03$ and $\omega=200$, and for
     general-relativistic models with $\Omega_b=0.3$ and
     $\Omega_b=0.032$.}
\label{fig:csw200bfit}
\end{center}
\end{figure}

The CMB power spectra are also affected by other
cosmological parameters, and it is possible that variation of
some other parameters might mimic the effect Brans-Dicke gravity.
For example, we plot the Brans-Dicke model with 
$\omega=200$ and $\Omega_{b}=0.030$
along with a general-relativity model with $\omega_{b}=0.032$
in Fig.~\ref{fig:csw200bfit}. This general-relativity model mimics the
Brans-Dicke model up to the first acoustic peak.
Note, however, that this different 
$\Omega_{b}$ model does not fit the polarization better---in
fact, the fit for the polarization is even worse.
Therefore, observation of the polarization may help to lift this 
degeneracy of parameters.

It may also be possible to mimic the Brans-Dicke model entirely
with a general-relativity model by adjusting more than one parameter. 
To investigate properly the possible degeneracy of the effect of 
varying $\omega$ with the possible effect of varying some
combination of other cosmological parameters,
we calculate the covariance matrix \cite{Fisher}.  This also
allows us to simultaneously estimate the precision with which
$\omega$ (actually $\omega^{-1}$) can be recovered with a CMB map.  
We first consider only a temperature map and later consider the
additional information that comes from the polarization.
If the true parameters which describe the Universe are given by
${\mathbf s}_0$, then the Fisher information matrix is defined by
\begin{equation}
\alpha_{ij}=\sum_{l} \frac{1}{\sigma_{l}^{2}} \left[\frac{\partial
C_{l}^{\rm TT}({\mathbf s}_0)}{\partial s_{i}}\frac{\partial
C_{l}^{\rm TT}({\mathbf s}_0)}{\partial s_{j}}\right].
\label{Fisher}
\end{equation} 
If the observed $C_l$'s are nearly
Gaussian distributed around $C_l({\mathbf s}_0)$ with variance
$\sigma_l$,
the covariance matrix $[{\mathcal C}]=[\alpha]^{-1}$ gives an estimate
of the standard errors that would be obtained from a maximum-likelihood
fit to the data. Approximately, the standard error with which
the parameter $s_i$ could be obtained (after marginalizing over
all other undetermined parameters) would be
$\sigma_{s_{i}}={\mathcal C}^{1/2}_{ii}$.

Consider a CMB experiment that maps the temperature of the
entire sky with a Gaussian beam of width $\theta_{\rm fwhm}$, 
with a noise per pixel of $\sigma_{\rm pix}$.  If a fraction
$f_{\rm sky}$ of the sky is used after a foreground cut, and the
noise in each pixel is uncorrelated, then the the standard error 
with which each $C_l^{\rm TT}$ can be recovered is
\begin{equation}
\sigma_{l}=\left[\frac{2}{(2l+1)f_{\rm sky}}\right]^{1/2} 
\left(C_{l}+w_{T}^{-1} e^{l^{2} \sigma_{b}^{2}}\right),
\end{equation}
where $\sigma_b = 7.42 \times 10^{-3} (\theta_{\rm
fwhm}/1^{\circ})$, and the inverse weight $w_{T}^{-1}$ is given
by
\begin{equation}
w_{T}^{-1} = 4\pi \sigma^{2}_{\rm pix} / N_{\rm pix},
\end{equation}
where $\sigma_{\rm pix}$ is the noise per pixel, and $ N_{\rm
pix}\simeq 40,000(\theta_{\rm fwhm})^{-2}$ is the number of
pixels.

\begin{figure}[htbp]
\begin{center}
\epsfig{file=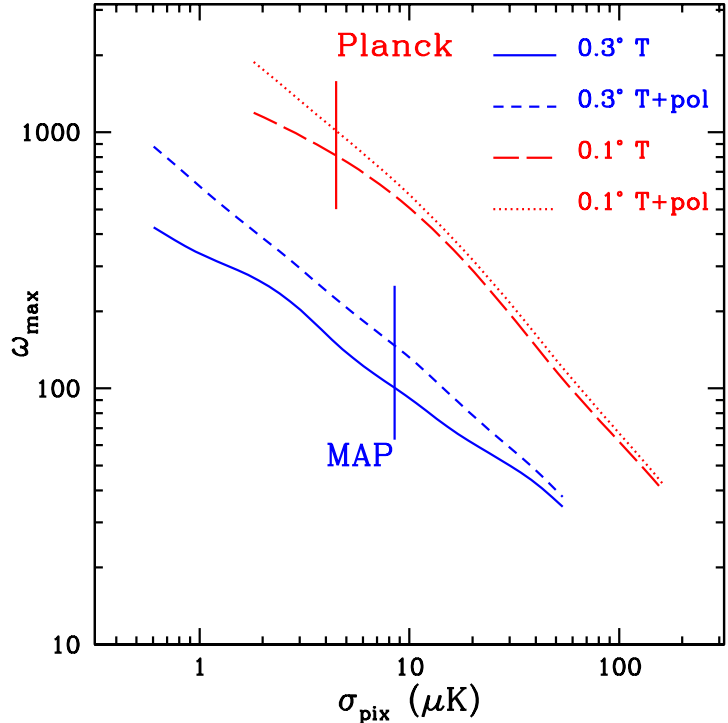,width=4in}
\caption{The largest finite value of $\omega$ that could be
     distinguished from infinity (i.e., general relativity) at the
     $2\sigma$ level as a function of the pixel noise of a given
     experiment that covers two-thirds of the sky.  The fiducial
     model here is a standard CDM model (no cosmological constant).
     We show results for two beamwidths, $\theta_{\rm
     fwhm}=0.1^\circ$ and  $\theta_{\rm fwhm}=0.3^\circ$.  We
     assume here that $\Omega_b$, $h$, $n_s$, $\tau$, and $Q$
     are marginalized over.  The solid curve corresponds to
     $\theta_{\rm fwhm}=0.3^{\circ}$ with temperature data only,
     and the short dashed curve also includes the polarization.
     The long dashed curve corresponds to $\theta_{\rm
     fwhm}=0.1^{\circ}$with temperature data only, and the
     dotted curve includes also the polarization.  The
     expected values of $\sigma_{\rm pix}$ for MAP and Planck
     are indicated.}
\label{Fisher_all}
\end{center}
\end{figure}

\begin{figure}[htbp]
\begin{center}
\epsfig{file=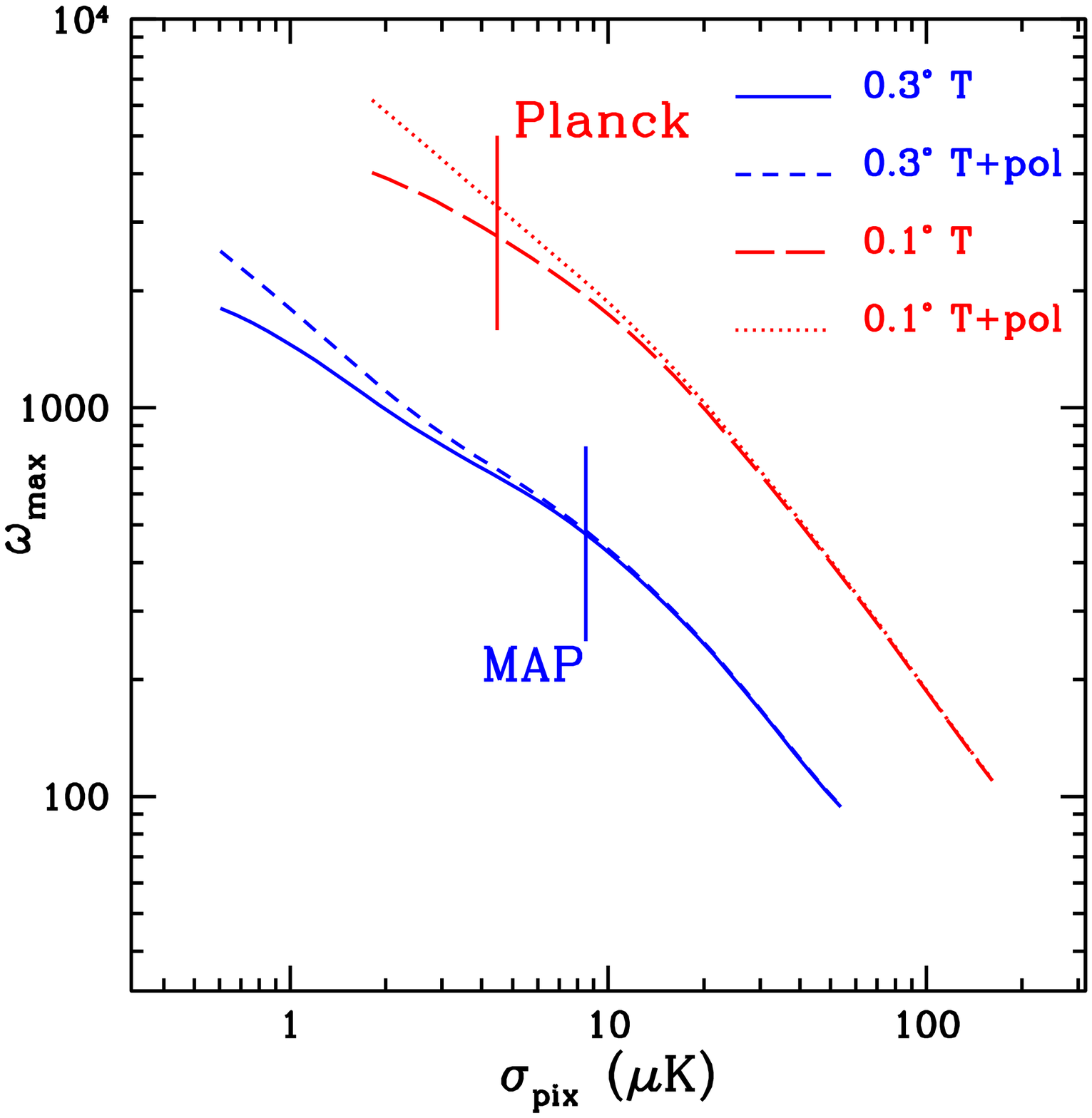,width=4in}
\caption{Same as Fig. \ref{Fisher_all}, except that all other
     parameters (except the normalization $Q$) are assumed to be
     known.}
\label{Fisher_Qonly}
\end{center}
\end{figure}

The goal of the MAP mission is to measure 
the temperature anisotropy with $\theta_{\rm fwhm}=0.3^{\circ}$ and 
$\sigma_{\rm pix}=20\, \mu {\rm K}$, which corresponds to 
$w_{T}^{-1}= 2\times 10^{-15}$ (assuming a one-year experiment).
The Planck Surveyor has a mission goal
of $\theta_{\rm fwhm}=0.1^{\circ}$ and $\sigma_{\rm pix}=5\, \mu{\rm K}$,
which corresponds to $w_{T}^{-1}= 6\times 10^{-17}$. Assuming $f_{\rm
sky}=0.67$, we calculate the Fisher information
matrix with
$\theta_{\rm fwhm}=0.3^{\circ}$ and $0.1^{\circ}$ for a variety of 
$w_{T}^{-1}$ values. The results are shown in
Figs.~\ref{Fisher_all} and \ref{Fisher_Qonly}.
Our fiducial model is a COBE-normalized flat CDM model with
$\{h_{0}, \Omega_{b}, n_{s}, \tau, 1/\omega, (Q/Q_{\rm COBE})^2\}
=\{0.65, 0.03, 1, 0.5, 0, 1\}$  where $n_{S}$ is the primordial
power-spectrum index, $\tau$ the reionization optical depth, and 
$Q_{\rm COBE}(n_s =1) = 
18 ~\mu {\rm K}$ is the COBE normalization \cite{COBE
Normalization}. We assume 3 generations of
massless neutrinos and no massive neutrinos, and consider scalar modes 
only. The derivatives of $C_{l}$ are calculated by varying each of 
the parameters by 0.5\%.  
To calculate the partial derivatives of the $C_l$'s with respect
to $1/\omega$, we compare the general-relativity
model with a Brans-Dicke model with $\omega=200$. Our calculation 
sums up modes up to $l\leq 3000$. We have checked to make sure that
our results are not sensitive to the step size for calculating
the derivative nor the cutoff of $l$.

\begin{table}
\begin{center}
\begin{tabular}{|l|l|l|l|l|l|l|}
 & $h_{0}$ & $\Omega_{b}$ & $n_{S}$ &
$\tau$ & $(Q/Q_{\rm COBE})^{2}$ &$1/\omega$\\
\hline
value & 0.65 & 0.03 & 1.0 & 0.5 & 1 &0\\
\hline
\hline
$\sigma_{\rm MAP} (T)$& 0.045 & 0.0054
& 0.043 & 0.057 & 0.057 & 0.0050\\
\hline
$\sigma_{\rm MAP} (T+P)$& 0.031 & 0.0036
& 0.031 & 0.035 & 0.053 & 0.0034\\
\hline
$\sigma_{\rm Planck}$& 0.0045
& 0.00049 & 0.0081 & 0.013 & 0.018 & 0.00062\\
\hline
$\sigma_{\rm Planck} (T+P)$& 0.0037
& 0.00040 & 0.0055 & 0.006 & 0.017 & 0.00049\\
\end{tabular}
\smallskip
\caption{Error estimates for parameters of an SCDM model. Here, MAP
     is assumed to have $\theta_{\rm fwhm}=0.3^{\circ}$ and
     $w^{-1}=2\times 10^{-15}$; Planck is assumed to have
     $\theta_{\rm fwhm}=0.1^{\circ}$ and $w^{-1}=6.3\times
     10^{-17}$.}
\label{fisher_scdm}
\end{center}
\end{table}

The value of $\sigma_{1/\omega}$ depends on the number and uncertainty
of other parameters.  Table \ref{fisher_scdm} lists the standard
errors that could be obtained by marginalizing over all others
for the various parameters we consider.  Fig.~\ref{Fisher_all}
plots the smallest value of $\omega$ that could be distinguished 
{}from $\omega=\infty$ (i.e., general relativity) at the 95\%
confidence level.  So, for example, if all the parameters listed
above were unknown and had to be determined from CMB data alone,
then the CMB would be marginally competitive with (current)
solar-system tests; i.e., if we marginalize over $Q_{\rm COBE}$,
$\Omega_b$, $h$, $n_s$, and $\tau$, then the smallest value of
$\omega$ that could be distinguished from $\infty$ is
$\simeq100$ for MAP and $\simeq 800$ for the Planck Surveyor. 

On the other hand, if we assume that all parameters except for
$\omega$ and the normalization can be determined completely 
{}from other experiments, then the sensitivity to a finite $\omega$
can be improved, as also illustrated in
Fig.~\ref{Fisher_Qonly}.  For example, a finite value of
$\omega$ as large as $\omega\simeq500$ could be detectable with
MAP and $\simeq 2500$ for Planck. 

The future satellite missions will measure not only the temperature
anisotropy but also the polarization. It is possible to improve the 
accuracy of cosmological-parameter determination by combining the 
temperature and polarization data, as Fig.~\ref{fig:csw200bfit}
illustrated that polarization may help break degeneracies in
parameter space.

To include the polarization data, we generalize Eq.~(\ref{Fisher}) to
\begin{equation}
\alpha_{ij}=\sum_{X,Y}\sum_{l} \left[\frac{\partial
C_{X,l}}{\partial s_{i}} \left[\Xi^{-1}\right]_{X,Y} \frac{\partial
C_{Y,l}}{\partial s_{j}}\right].
\label{FisherPol}
\end{equation}
Here, $X, Y=$TT, GG, CC, TG, and $[\Xi^{-1}]_{X,Y}$ are elements of
the inverse noise covariance matrix $\Xi$. The elements of $\Xi$ were 
given in Refs. \cite{CMB polarization KKS,CMB polarization SZ}.
If the two linear-polarization states are given equal integration
times, the total number of photons available for temperature
measurement are twice of that for polarization measurement, thus
\begin{equation}
(\sigma_{\rm pix}^{T})^{2}=\frac{1}{2}(\sigma^{P}_{\rm pix})^2.
\end{equation}
If the number of pixels are equal, then 
\begin{equation}
w_{T}^{-1}=w_{P}^{-1}.
\label{wt/wp}
\end{equation}

The results obtained from combining the temperature and
polarization data are also plotted in Figs.~\ref{Fisher_all} and 
\ref{Fisher_Qonly}, and these plots show that the sensitivity to
the Brans-Dicke parameter $\omega$ could be 
further improved by including polarization. The effect is
particularly strong when the
effect of Brans-Dicke gravity is degenerate with that of
variation of other cosmological parameters in a temperature map. 
By including the polarization data in MAP,
the CMB should be sensitive to models with $\omega<150$ at 95\%
CL when all other parameters are undetermined, or $\omega<500$ when
only CMB normalization is undetermined (in this case, there is not
much gain from polarization). For Planck, these numbers
are $\omega<1000$ and $\omega<3200$, respectively.  On the other
hand, if all the other parameters are known, the benefit gained
{}from adding polarization measurement is less obvious; only for
detectors with pixel noise less than $5\mu$K there is a difference.

\begin{figure}[htbp]
\begin{center}
\epsfig{file=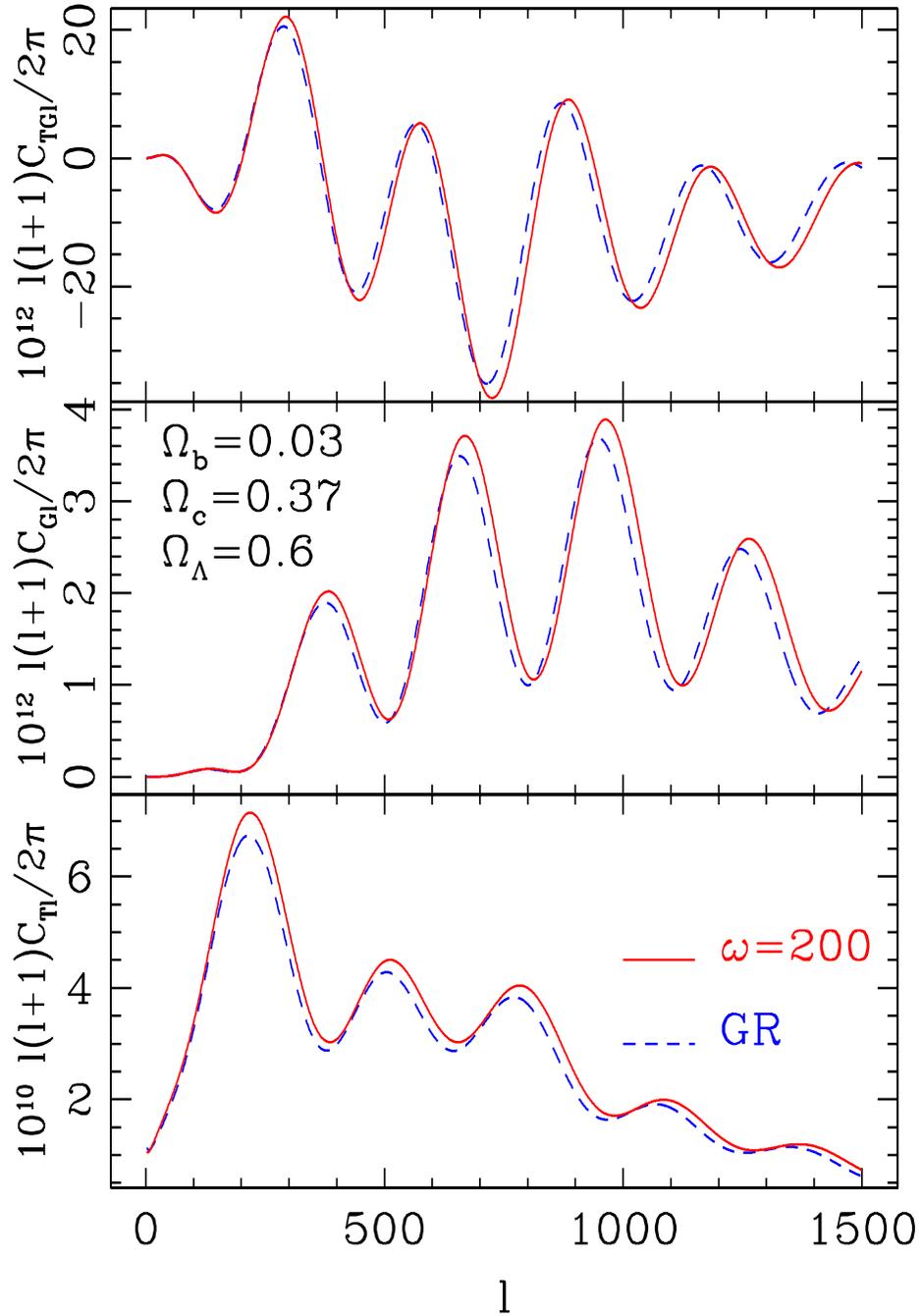,width=5in,height=7.5in}
\caption{CMB power spectra for $\Lambda$CDM models in
       Brans-Dicke theory with $\omega=200$ and in general relativity.}
\label{fig:lw200}
\end{center}
\end{figure}

\begin{figure}[htbp]
\begin{center}
\epsfig{file=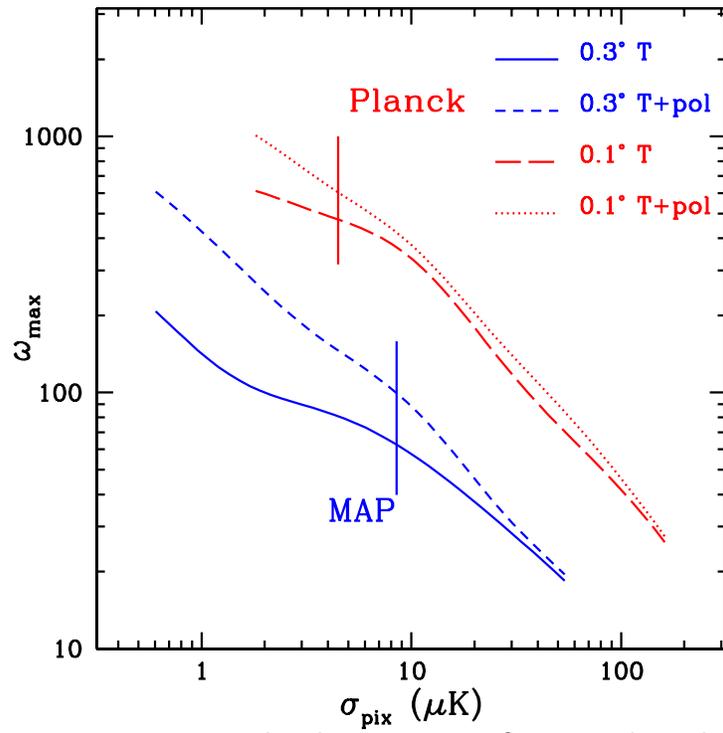,width=4in}
\caption{The same as Fig. \ref{Fisher_all}, except that here we
use a flat cosmological-constant model with a
nonrelativistic-matter density $\Omega_0=0.4$.}
\label{Fisher_all_L}
\end{center}
\end{figure}

\begin{figure}[htbp]
\begin{center}
\epsfig{file=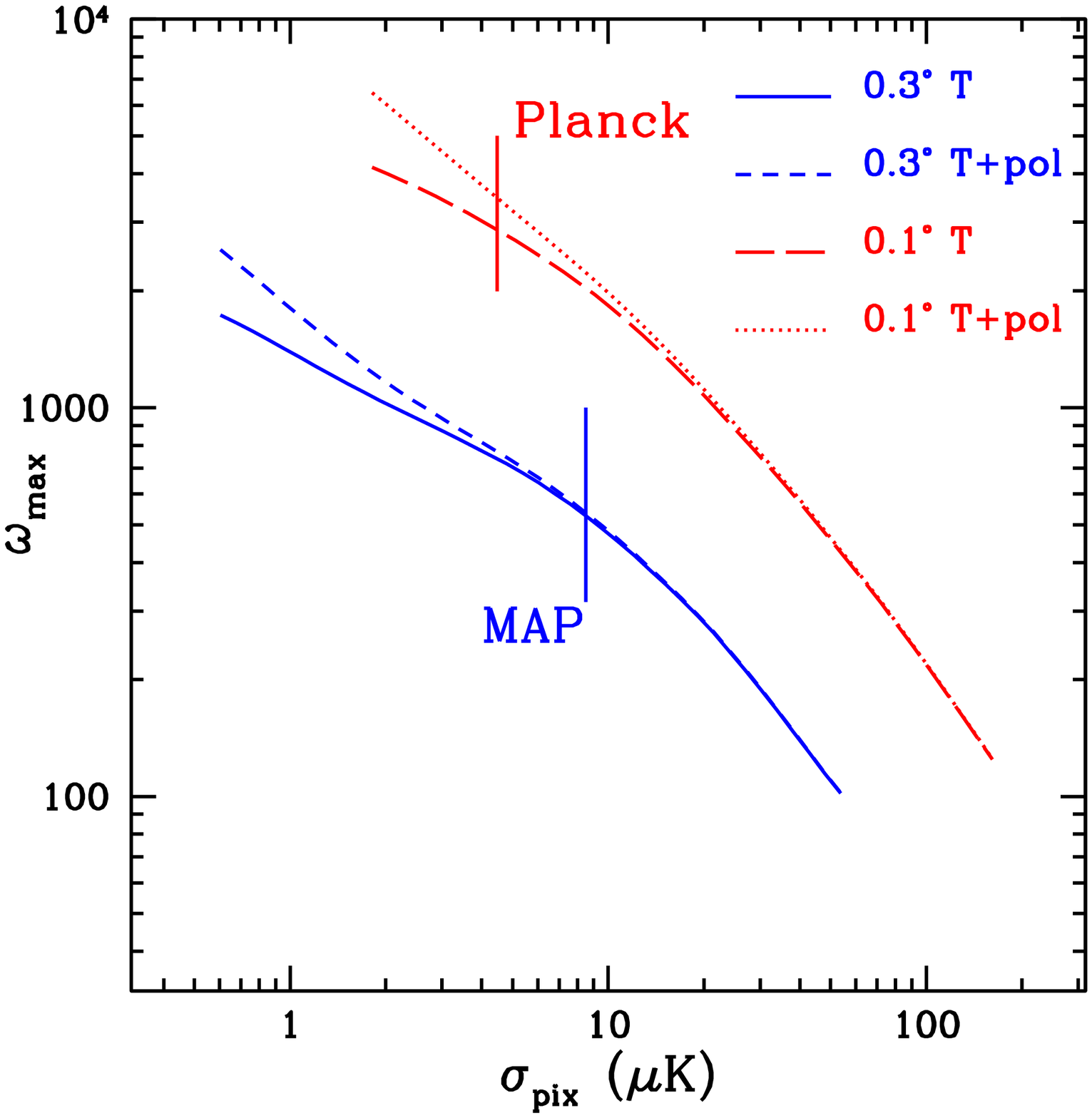,width=4in}
\caption{Same as Fig. \ref{Fisher_all_L}, except that all other
     parameters (except the normalization $Q$) are assumed to be
     known.} 
\label{Fisher_Qonly_L}
\end{center}
\end{figure}

There has been much recent interest in models with a nonzero
cosmological constant prompted, in particular, by the evidence
for an accelerating universe from supernovae \cite{supernovae}.
We have also performed our calculations for a $\Lambda$CDM model,
and the results are shown in Fig. \ref{fig:lw200} for
$\omega=200$.  The results for the Fisher-matrix analysis are
shown in Figs.~\ref{Fisher_all_L} and \ref{Fisher_Qonly_L}, and
they are similar to those in the SCDM case.

Finally, we have noted before, if $\omega$ is finite, then
$\Omega$ is not precisely equal to unity [cf.,
Eq. (\ref{eq:changedOmega})], and one might wonder whether the
effects on the CMB power spectra of varying $\omega$ can be
mimicked by varying $\Omega$.  We have checked that for the
Brans-Dicke models we have investigated, the change in the CMB
power spectra is much too large to be attributed to this shift
in $\Omega$.

\section{Conclusion and Summary}

We have developed a formalism for calculating the CMB anisotropy in
cosmological models with Brans-Dicke gravity. This was done by modifying
standard Boltzmann codes for CMB power spectra. Because Brans-Dicke
theory satisfies the medium-strong equivalence principle, only those
equations determining the evolution of the metric need to be modified; the 
equations of motion for various matter or radiation components are the
same as in general relativity. One boundary condition for the
Brans-Dicke theory
is determined by requiring $\phi=\frac{2\omega+4}{2\omega+3}$ at the
current epoch. Another is given by adopting the
Brans-Dicke initial condition $a^{2} \phi'=0$ at early time
(after the annihilation of electron-positron pairs).

This formalism is then used to calculate the CMB power spectra
in several models.  We find that in Brans-Dicke models,
both the height and width of the acoustic peaks are changed. 
While there is some degeneracy with different cosmological
parameters at low $l$ in a temperature map, we demonstrate that
the effect can be distinguished by going to higher acoustic
peaks and by observing the polarization of the CMB. 
Our results show that with high-quality CMB data,  the CMB
anisotropy may provide a powerful test for Brans-Dicke theory
that is competitive (and complementary) to solar-system tests. 

As an example, we examined a flat SCDM model. MAP temperature
data should be able to distinguish Brans-Dicke gravity with
$\omega<100$ from general relativity at the 95\% CL 
if all other parameters must be simultaneously determined from
the CMB, or $\omega<500$ if all other
parameters except for the CMB normalization are fixed. With
Planck, these numbers are $800$ and $2500$ respectively.

Furthermore, even better results are achievable if both temperature
and polarization data are used. For MAP, the two limits are
raised to $150$ and $800$, respectively, and for Planck, $1000$
and $3200$, respectively.  We also examined the case of a flat
$\Lambda$CDM model and found similar results.

In conclusion, the differences between the CMB power spectra
expected in general relativity and those in Brans-Dicke models
with acceptable values of $\omega$ are small.  However, our
Fisher-matrix analysis shows that if systematic effects can be
controlled, then the CMB sensitivity (from Planck) to a finite
value of $\omega$ might be competitive with that from
solar-system tests.  We re-emphasize that the CMB will
provide a new and independent test of gravity in stronger fields 
and at earlier times.  Thus, it is conceivable that the CMB will
provide a unique test of some scalar-tensor theories in which
$\omega$ would have been smaller at earlier times.

\acknowledgments

We thank A. Liddle and A. Mazumdar for helpful
discussion.  This work was supported by D.O.E. Contract
No. DEFG02-92-ER 40699, NASA NAG5-3091, and the Alfred P. Sloan
Foundation.

\appendix
\section{Numerical Implementation}

In this Appendix we briefly describe the numerical implementation of
the calculation. 
First we consider the background evolution. The boundary conditions
for the Brans-Dicke field $\phi$ are given in
Eqs.~(\ref{init1}) and (\ref{init2}). From Eq.~(\ref{init1}), we
know the end-point value of  
$\phi$ but not the initial value.  We pick an arbitrary
epoch with temperature 10 eV$\ll T \ll 0.5$ MeV, then pick some value
of $\phi$ with $\phi'=0$ and integrate forward until the scale
factor $a=1$.\footnote{In principle, one may also integrate backwards from
the present epoch to this
early stage and thus obtain the ``initial value'', but this procedure
is susceptible to numerical instability \cite{BBN recent}.} This
process is reiterated with different trial initial values of $\phi$ until
the condition $\phi_{0}=(2\omega+4)/(2\omega+3)$ is satisfied to the 
required precision. Formally, the process is equivalent to numerical 
root finding, and we use a Brent algorithm \cite{recipe} to find the root.

In the original {\tt CMBFAST} code, any epoch in the evolution
of the Universe is
specified by the cosmic scale factor $a$, and the time $\tau$ is
obtained by integrating Eq.~(\ref{daoa}).
In Brans-Dicke theory, the 
value of $a'/a$ is given by Eq. (\ref{modified
FRW}), and it is no longer convenient to use $a$ as the
argument. Instead, we use
$\tau$ to specify the cosmic time, and the scale
factor $a$ is obtained by solving the whole set of
background-evolution differential equations. This is done in the
beginning of the calculation, and the values of 
$\{\tau, a(\tau), a'(\tau), \phi(\tau), \phi'(\tau)\}$ are
then stored. Subsequently, given either $\tau$ or $a$, the whole set of these
values corresponding to that epoch can be obtained by 
lookup and/or interpolation. The second-order derivatives $a''$
and $\phi''$ can also be obtained by numerical difference.

This more complicated implementation of the cosmic history also
demands modification in other parts of the code. 
In the original code, the expansion rate is calculated from the density
of the universe using the Friedman equation; in this new code,
we replace it by interpolation from the stored data  in all
such cases. 

For example, in the original code,  
the baryon temperature, ionization fraction, and baryon sound speed are
given as a functions
of $a$, and are calculated by a simple integration of the 
Friedmann equation. In the new scheme, this calculation is modified so that
the value of $a'$ is obtained by lookup or/and interpolation of the stored
data. We also modified the code for the calculation of the baryon sound 
speed. In the original calculation, there are occasional jumps in the
baryon sound speed, which may be due to the truncation error in calculating
the derivative ${\rm d} T_{b}/{\rm d} \ln a$. We have modified
the algorithm so that there is no jump in this calculation.  However, our
tests show that these occasional jumps do not have any significant
effect on the end result, probably because only interpolated values
are used, and the result is mostly important only in a limited range.

For the calculation of the perturbative part, we note that in
Brans-Dicke theory, the equation of motion for the matter or radiation
are the same as in general relativity. All we need to do is to replace
the perturbed Einstein equations by
Eqs.~(\ref{perturbed})--(\ref{perturbed-last}). 
Following the original code, we use Eq.~(\ref{forcecon}) to force
conservation of energy and reduce the numerical error in solving
the ordinary differential equations. We have tested that when
$\omega \to \infty$  we recover the general-relativistic result
produced by the standard code.

Finally, the temperature and polarization anisotropy may be obtained
by integrating Eqs.~(\ref{losint1})--(\ref{losint2}). The expression
for the source function is the same, but note that there are
Brans-Dicke corrections to the derivatives of 
metric perturbations as given in
Eqs.~(\ref{alpha1})--(\ref{alpha2}), and these must be
implemented in the code.

\end{document}